\newtheorem{theorem}{Theorem}
\newtheorem{lemma}{Lemma}
\newtheorem{proposition}{Proposition}

\documentclass[german]{article}
\usepackage{graphics}
\usepackage{amssymb,amsmath}
\newcommand{\T}{\textstyle}
\newcommand{\hyp}[1]{\mathbb{H}^{#1}}
\newcommand{\ul}[1]{\underline{#1}}
\newcommand{\bn}[1]{|{#1}|}
\newcommand{\norm}[1]{\|#1\|}
\newcommand{\R}[1]{\mathbb{R}^{#1}}
\begin{document}
\title{AdS/CFT correspondence in the Euclidean context}

\author{Hanno Gottschalk and Horst Thaler\\ 
%
Institut f\"ur angewandte Mathematik, Universit\"at Bonn, Germany \\
e-mail: gottscha@wiener.iam.uni-bonn.de \and Dipartimento di Matematica e Informatica, Universit\`{a} di Camerino, Italy \\
e-mail: horst.thaler@unicam.it}
%
%
%
\maketitle
\begin{abstract}
We study two possible prescriptions for the AdS/CFT correspondence by means of functional integrals. The considerations are non-perturbative and reveal certain divergencies which
turn out to be harmless, in the sense that reflection positivity and conformal invariance are not destroyed.
\end{abstract}
\section{Introduction}
\label{intro} In this article we investigate the AdS/CFT
correspondence for scalar fields within the Euclidean approach.
Originally, this conjecture was formulated within the string
theoretic context \cite{Ma}. Soon afterwards it was discovered that
it makes perfect sense in a purely quantum field theoretic setting
\cite{Wi}. This conjecture states that a quantum field theory (QFT)
on AdS space gives rise to a conformal QFT (CFT) on its boundary and vice
versa. Within the algebraic approach to QFT this correspondence can
be made precise. The idea is to identify algebras of observables in wedge-like
regions on AdS space with corresponding algebras in double cones on
the boundary, see \cite{Rea}. We hope that this work can contribute
to the recent discussion on the mathematical status of non-algebraic
AdS/CFT.

We are interested in the passage from AdS-QFT to CFT by means of
functional integrals. Without taking recourse to perturbative
arguments we succeed in constructing functional integrals within the
infinite dimensional setting. The Euclidean field theory of an
interacting QFT is described through a probability measure
$d\mu=e^{-V}d\mu_C/\int e^{-V}d\mu_C$, defined on an appropriate
distribution space on the Riemannian counterparts of AdS spaces,
which are hyperbolic spaces. The Gaussian measure $d\mu_C$ with
covariance $C$ specifies the underlying free theory and the density
$e^{-V}$ accounts for the interaction. The measure $d\mu$ should
satisfy the Osterwalder-Schrader axioms in order to make a passage
from hyperbolic to AdS-spaces possible \cite{BEM,JR}.

On hyperbolic spaces there are two choices
of invariant covariance operators, denoted $G_\pm$, since there are two linearly independent fundamental solutions to the
equation
$$(-\Delta+m^2)G(\ul{x},\ul{x}')=\delta(\ul{x},\ul{x}').$$
This follows from the fact that, due to invariance, $G$ has to be a
function of the geodesic distance $d(\ul{x},\ul{x}')$, therefore the
resulting equation for $G(d)$ involves only the radial part of the
Laplacian which can be transformed to a hypergeometric equation
possessing two linearly independent solutions. This work is inspired
by the ideas in \cite{DuRe} (see also \cite{Reb}) where two natural
prescriptions for the AdS/CFT correspondence are compared and shown
to essentially agree. One way is to define a Laplace transform where
the source term is restricted to the boundary, i.e.
\begin{equation}\label{gen2}
\tilde{Z}(f)/\tilde{Z}(0)=\int e^{-V(\phi)}
e^{\partial\phi(f)}d\mu_C(\phi)/\tilde{Z}(0).
\end{equation}
At this place $\partial\phi$ means the restriction
of the bulk field to the boundary. Below we shall see how to make this definition rigorous using a proper scaling.
It turns out that in general nontrivial results for (\ref{gen2}) can be obtained only
through the multiplication with a regularizing factor which nonetheless doesn't destroy reflection positivity and conformal invariance.
Another possibility is to fix the values of the bulk field on the boundary by insertion of a
delta function, so that heuristically we set
\begin{equation}\label{gen1}
Z(f)/Z(0)=\int e^{-V(\phi)}\delta(\partial\phi-f)d\mu_C(\phi)/Z(0).
\end{equation}
It will turn out that the correct choice for $C$ is to take $G_+$ in case (\ref{gen2}) and
$G_-$ for (\ref{gen1}). Essentially, the definition of (\ref{gen1}) rests on the splitting of $G_-$ into a ``bulk-part" and a ``boundary-part". For a related discussion about quantum fields on manifolds with a boundary, look at \cite{Ha}. Another viewpoint on the relation between bulk and boundary fields, using representation theoretic arguments, can be found in \cite{Do}. The construction makes it also explicit that the two functionals agree up to the multiplication of test functions with a constant factor when both are defined.

In section 2 we introduce various propagators which serve as
building blocks for the functional integrals, in particular the
splitting of $G_-$ is proven. In section 3 we show how to give a
rigorous meaning to expressions (\ref{gen2}) and (\ref{gen1}). Then
in section 4 we treat $P(\phi)_2$ models for concreteness. In
section 5 we go over to discuss the two basic axiomatic properties
of reflection positivity and conformal invariance.\newpage
\section{Propagators on the hyperbolic space}
There are various propagators needed for the definition of AdS/CFT functional integrals, which we introduce in this section. Let us consider the upper half-space model of the $(d+1)$-dimensional hyperbolic space
$\hyp{d+1}:=\{(z,x)\in \mathbb{R}^{d+1}: z>0\},$
equipped with the Riemannian metric $1/z^2(dz^2+dx_1^2+\cdots+dx_{d}^2)$.
The Green's functions $G_\pm$ are explicitly given by
\begin{equation}\label{green}
G_\pm(z,x;z',x')=\gamma_{\pm}(2u)^{-\Delta_{\pm}}F(\Delta_{\pm},\Delta_{\pm}+{\T \frac{1-d}{2}};2\Delta_{\pm}
+1-d;-2u^{-1})
\end{equation}
where $u=\frac{(z-z')^2+(x-x')^2}{2zz'}$, $\Delta_{\pm}=\frac{d}{2}\pm\frac{1}{2}
\sqrt{d^2+4m^2}=:\frac{d}{2}\pm\nu, \nu> 0$ and $\gamma_\pm =\frac{\Gamma(\Delta_\pm)}{2\pi^{d/2}\Gamma(\Delta_\pm+1-\frac{d}{2})}$. $F$ is the hypergeometric function which for $\zeta\in \mathbb{C}$ with $|\zeta|<1$ is given
by the absolutely convergent series
\begin{equation}\label{hypergeo}
F(a,b;c;\zeta)=1+\frac{ab}{c}\zeta+\frac{a(a+1)b(b+1)}{c(c+1)}\zeta^2+\cdots
\end{equation}
Its analytic continuation to $\mathbb{C}\backslash [1,\infty)$ is given by the integral representation
$$
F(a,b;c;\zeta)=\frac{\Gamma(c)}{\Gamma(b)\Gamma(c-b)}\int_0^1 t^{b-1}(1-t)^{c-b-1}(1-\zeta t)^{-a}dt, \quad {\rm if \hspace{1ex} Re}c>{\rm Re}b>0.
$$
It should be noted that $G_+$ is the integral kernel of the inverse $(-\Delta+m^2)^{-1}$ in
$L^2(\hyp{d+1})$.

We would like to obtain a conformal theory on the boundary at infinity $(z\rightarrow 0)$. On the level of propagators this is achieved by taking appropriate scaled limits.
From (\ref{green}) we get as pointwise limits the bulk-to-boundary propagators
$$
H_\pm(z,x;x')=\lim_{z'\rightarrow 0}z'^{-\Delta_\pm}G_\pm(z,x;z',x')
=\gamma_\pm\left(\frac{z}{z^2+(x-x')^2}\right)^{\Delta_\pm}
$$
and the boundary propagators
\begin{equation}\label{alpha}
\alpha_\pm(x,x')=\lim_{z \rightarrow 0 }z^{-\Delta_\pm}H_\pm(z,x;x')
=\gamma_\pm(x-x')^{-2\Delta_\pm}.
\end{equation}
Since $2\Delta_+ \geq d$, the kernels $\alpha_+$ have a non-integrable singularity. They will be understood to be regularized by analytic continuation to values $\nu \neq 0,1,2,\ldots$, see \cite{GeSh}. Hence, whenever $\alpha_+$ is involved in some argument, statements will hold
with the exception of singular points.\\
{\it Notation.}
The Fourier transform is defined as $\hat{f}(k)=1/(2\pi)^{d/2}\int_{\mathbb{R}^d}f(x)^{-ikx}dx$. We use the notation
$|\zeta|$ for the absolute value of a complex number $\zeta$, as well as $|k|$ for the Euclidean norm of a vector $k\in \R{d}$. Tuples $(z,x)$ will also be denoted by $\underline{x}$.

Then the Fourier transforms of $H_{\pm}(z,x;x')$ and $\alpha_{\pm}(0,x')$ with respect to $x'\in \mathbb{R}^d$ read
\begin{equation}\label{fourierh}
\hat{H}_{\pm}(\underline{x},k)=\frac{1}{(2\pi)^{\frac{d}{2}}\Gamma(1\pm\nu)}e^{ikx}\left(\frac{\bn{k}}{2}\right)^{\pm \nu}z^{\frac{d}{2}}K_{\nu}(\bn{k}z),
\end{equation}
and
\begin{equation}\label{fourierk}
\hat{\alpha}_{\pm}(k)=\frac{\Gamma(\mp\nu)}{2(2\pi)^{\frac{d}{2}}\Gamma(1\pm\nu)}\left(\frac{\bn{k}}{2}\right)^{\pm 2\nu}=:C_{-\nu}\left(\frac{\bn{k}}{2}\right)^{\pm 2\nu},
\end{equation}
where $K_\nu$ is the modified Bessel function of the second kind which is given by
$$
K_\nu(\zeta)=\frac{1}{2}\left(\frac{\zeta}{2}\right)^\nu \int_{-\infty}^\infty \frac{e^{-t-\zeta^2/4t}}{t^{\nu+1}}dt, \quad |{\rm arg}\zeta|<\frac{\pi}{2},\;{\rm Re}\zeta^2>0.
$$
For small arguments it behaves like
$K_\nu(\zeta)\sim \frac{1}{2}\Gamma(\nu)\left(\frac{\zeta}{2}\right)^{-\nu},\;\nu>0$.
\begin{lemma}\label{lemmasplitt}
With $c:=2\nu$ we have
\begin{equation}\label{splitting}
G_-(\ul{x},\ul{x}')= G_+(\ul{x},\ul{x}') + \int_{\mathbb{R}^d}\int_{\mathbb{R}^d}
 H_+(\ul{x},y)c^2\alpha_-(y,y')H_+(\ul{x}',y')dy dy'.
\end{equation}
\end{lemma}
{\it Proof.} In \cite{muwi} it was shown that
\begin{equation}
G_-(\ul{x},\ul{x}')= G_+(\ul{x},\ul{x}') + c\int_{\mathbb{R}^d}
 H_+(\ul{x},y)H_-(\ul{x}',y)dy.
\end{equation}
Let $\alpha_-(\,\cdot\,)$ be the function $y'\rightarrow \alpha_-(0,y')$ then for $\nu > 0$
\begin{equation}\label{proofsplitt1}
\int_{\mathbb{R}^d}\alpha_-(y,y')H_+(\ul{x}',y')dy'
=(\alpha_{-}(\,\cdot\,)\ast H_{+}(z',x';\,\cdot\,))(y),
\end{equation}
where $\ast$ means convolution.
Therefore, for $2\nu <d$,
$$
(\alpha_{-}(\,\cdot\,)\ast H_{+}(z',x';\,\cdot\,))(y)=\int_{\mathbb{R}^d}e^{iky}\hat{\alpha}_{-}(k)\hat{H}_{+}(z',x';k)dk
$$
\begin{equation}\label{before}
=\frac{1}{c (2\pi)^{d}\Gamma(1-\nu)}\int_{\mathbb{R}^d}e^{ik(x'+y)}\left(\frac{\bn{k}}{2}\right)^{- \nu}z'^{\frac{d}{2}}K_{\nu}(\bn{k}z')dk.
\end{equation}
On the other hand, for $2\nu <d$,
\begin{equation}\label{before1}
\frac{1}{c}H_{-}(\underline{x}',y)=\frac{1}{c(2\pi)^{\frac{d}{2}}}\int_{\mathbb{R}^d}e^{iky}\hat{H}_{-}(\underline{x}',k)dk,
\end{equation}
Using Morera's theorem \cite{BeGa}, it is not difficult to see that for fixed $\underline{x}',y$ the left-hand sides of (\ref{before}) and (\ref{before1}) are holomorphic as functions of the parameter $\nu >0$ and, because they agree for $2\nu <d$, the result follows. $\qquad \Box$ \\
{\it Remark 1.} Equation (\ref{splitting}) presents the splitting of $G_-$ into a ``bulk-part" and a ``boundary-part". Although it is a covariance on $\hyp{d+1}$, the ``boundary-part" is named like this because it contains the boundary covariance $\alpha_-$. We note
that the ``bulk-part" vanishes with respect to the scaling $z^{-\Delta_-}$. Moreover, a splitting for $G_+$ like that in Lemma 1 into a sum of two covariances is not possible. In order to get the right boundary covariance $\alpha_+(x,x')=\lim_{z\rightarrow 0}z^{-2\Delta_+}G_+(z,x;z,x')$ the ``bulk-part"
should scale like $z^a$, in any argument, with $a>\Delta_+$ in order to vanish
with respect to the scaling $z^{-\Delta_+}$, but such a covariance is not available among the solutions of $(-\Delta+m^2)f=\delta\,(=0)$.

\section{Construction and definition of functional integrals}
First we try to give a meaning to the functional integral (\ref{gen1}).
For $2\nu <d$, $\alpha_-$ is a positive covariance and in this parameter range
the splitting given in $(\ref{splitting})$
entails the corresponding splitting for the random fields,
$$\phi_-(\ul{x})=\phi_+(\ul{x})+c H_+\phi_\alpha(\ul{x}),$$
where $H_+\phi_\alpha(\ul{x}) := \int_{\mathbb{R}^d}H_+(\ul{x},y)\phi_\alpha(y)dy,$
and $\phi_-,\, \phi_+,\phi_\alpha $ are the Gaussian random fields with covariances
$G_-,\,G_+$ and $\alpha_-$ respectively.
More precisely, $\phi_+,\phi_{\alpha}$ have to be understood as the first and second component of the following product measure space
$$({\cal D}(\hyp{d+1})'\times{\cal S}(\mathbb{R}^d)',
{\cal B}({\cal D}(\hyp{d+1})')\otimes{\cal B}({\cal S}(\mathbb{R}^d)'),\mu_{G_+}\otimes\mu_{\alpha_{-}}).$$
${\cal D}(\hyp{d+1})$ stands for the space of infinitely differentiable real-valued functions with compact support on $\hyp{d+1}$ and ${\cal S}(\R{d})$ denotes the Schwartz space of rapidly decreasing real-valued functions on $\R{d}$. The primes indicate the topological duals, or
distribution spaces.
Finally, ${\cal B}$ stands for the Borel $\sigma$-algebras obtained from the respective weak-$\ast$ topologies. Then we have
$$\mathbb{E}_{\mu_{G_+}\otimes\mu_{\alpha_-}}
[(\phi_+(f)+c H_+\phi_\alpha(f))(\phi_+(g)+c H_+\phi_\alpha(g))]$$
$$=\mathbb{E}_{\mu_{G_+}}[\phi_+(f)\phi_+(g)] +
c^2\mathbb{E}_{\mu_{\alpha_-}}[H_+\phi_{\alpha}(f)H_+\phi_{\alpha}(g)],$$
because the other terms vanish due to the product measure and the fact that
the expectations of the fields vanish.
But the last line is just the splitting (\ref{splitting}).
For this reason we may write for $\nu < \frac{d}{2}$
\begin{equation}\label{prodint}
\int_{{\cal D}'} F(\phi_-)d\mu_{G_{-}}(\phi_-) = \int_{{\cal D}'\times{\cal S}'} F(\phi_+ +
c H_+\phi_{\alpha})d(\mu_{G_+}\otimes\mu_{\alpha})(\phi_+,\phi_{\alpha}).
\end{equation}
So far, $F$ can be a general integrable function. Usually one considers the form $F=e^{-V}$ with $V$ a local potential (with or without cut-offs). \\
{\it Remark 2.} The bound $-\nu >-\frac{d}{2}$ for the field $\phi_-$ is dictated by the positivity of $\alpha_-$.
For $d=1$ this is larger than the unitary bound $-\nu >-1$ (in our notation). The same bound is needed when $\alpha_-$ is asked to be reflection positive, see \cite[Theorem 6.2.4]{GlJa}. For $d\geq 2$ reflection positivity imposes
the usual unitary bound $-\nu>-1.$

In order to cope with the delta function we shall boil down things to a finite dimensional
approximation for the boundary field, insert the delta function in this case, perform integration
over the (finite-dimensional) boundary field and then remove the approximation again.
This is done in two steps. \newline
{\it Step 1.} We approximate the boundary covariance operator $\alpha_{-}$ by
covariance operators
which possess bounded inverses in $L^2(\mathbb{R}^d)$. First we note that
\begin{equation}\label{alfourier}
\int_{\mathbb{R}^d}\int_{\mathbb{R}^d}f(x)\alpha_{-}(x,y)f(y)dxdy =
C_{-\nu}\int_{\mathbb{R}^d} \bn{k}^{-2\nu}|\hat{f}|^2dk, \quad f\in {\cal S}(\R{d}).
\end{equation}
From (\ref{alfourier}) we see that the bounded approximations can be defined as follows
$$
(f,\alpha_{-}^nf):=C_{- \nu}\int_{\mathbb{R}^d}\chi_n(\bn{k}) |\hat{f}|^2dk,\quad (n\in\mathbb{N}),
$$
where
$$\chi_n(\bn{k}):=\left\{
\begin{array}{ll}
n^{2\nu}, &  {\rm for}\hspace{1ex} \bn{k}\leq \frac{1}{n},\\
\bn{k}^{-2\nu}, & {\rm for}\hspace{1ex} \frac{1}{n}<\bn{k}\leq n,\\
n^{-2\nu}, & {\rm for}\hspace{1ex} \bn{k} > n.
\end{array}\right.
$$
Obviously, for their inverses we obtain
$$
(f,(\alpha_{-}^n)^{-1}f)=(C_{-\nu})^{-1}\int_{\mathbb{R}^d}(\chi_n(\bn{k}))^{-1} |\hat{f}|^2dk.
$$
\newline
{\it Step 2.} Next we consider finite dimensional approximations for the boundary field
$\phi_{\alpha^n}$ ($n$ arbitrary) with covariance $\alpha_{-}^n$. This approximation is performed by
$p_m \phi_{\alpha^n}$, where
$p_m$ is the projection on the subspace spanned by the first $m$ basis
elements of a Hilbert space basis $(e_i)_{i\geq 1}$ of $L^2(\mathbb{R}^d)$. In order that the matrix elements $(e_i,\alpha_-^n e_j)$ be defined, we choose the basis elements to be Schwartz functions, which is possible, since Schwartz spaces are separable.
Making in addition the identification
$\eta : p_m\phi_{\alpha} \rightarrow \psi_{\alpha}=
((\phi_\alpha)(e_1),\ldots,(\phi_\alpha)(e_m))^t\in \mathbb{R}^m$,
we see that the integral (\ref{prodint}) takes the form
$$
C_{A_{-}}\int_{\mathbb{R}^m} \int_{{\cal D}'} F(\phi_+ + c H_+(\eta^{-1}\psi_{\alpha}))
d\mu_{G_+}(\phi_+)e^{-\frac{1}{2}(\psi_{\alpha},A_-\psi_{\alpha})}d\psi_{\alpha}
,$$
where $A_-:=(\eta p_m \alpha_{-}^n p_m \eta^{-1})^{-1}=\eta (p_m \alpha_{-}^n p_m)^{-1} \eta^{-1}$ and $C_{A_-}=\frac{|{\rm det}A_-|^{\frac{1}{2}}}{(2\pi)^{\frac{d}{2}}}.$
Now it is possible to insert the delta function and we get
$$
C_{A_{-}}\int_{\mathbb{R}^m} \int_{{\cal D}'} \delta(\psi_{\alpha}-\eta p_m f)F(\phi_+ +
c H_+(\eta^{-1}\psi_{\alpha}))d\mu_{G_+}(\phi_+)
e^{-\frac{1}{2}(\psi_{\alpha},A_-\psi_{\alpha})}d\psi_{\alpha}
$$
\begin{equation}\label{deltaint}
=C_{A_{-}}e^{-\frac{1}{2}(f,(p_m\alpha_-^n p_m)^{-1}f)}\int_{{\cal D}'} F(\phi_+ +
c H_+(p_m f))d\mu_{G_+}(\phi_+)=:Z_{m,n}(f).
\end{equation}
We notice that in the quotient $Z_{m,n}(f)/Z_{m,n}(0)$ the constant $C_{A_-}$ drops out.
The uniform convergence $p_m \rightarrow 1$ leads to $(p_m\alpha_{-}^n p_m)^{-1}f
\rightarrow (\alpha_{-}^{n})^{-1}f$, due to the boundedness of operators, see \cite[Theorem 5.11]{We}. Let $H_+(z;\,\cdot\,)$ denote the function $x\rightarrow H(z,x;0)$ then we have $\norm{(H_+p_mf)(z,\,\cdot\,)}_2=\norm{H_+(z;\,\cdot\,)\ast p_mf}_2\leq \norm{H_+(z;\,\cdot\,)}_1\norm{p_mf}_2$ by Young's inequality. Moreover, $z \rightarrow \norm{H_+(z;\,\cdot\,)}_1$ remains bounded if $z$ varies in a bounded subset of $(0,\infty)$. Therefore, under the assumption that
\begin{equation}\label{assumption1}
\norm{F(\,\cdot\,+c H_+(p_mf)-F(\,\cdot\,+c H_+(p_nf))}_{L^1(\mu_{G_+})}\leq {\rm const}\norm{c(H_+(p_mf-p_nf))|_\Lambda}_2,
\end{equation}
with a bounded $\Lambda\subset \hyp{d}$ we get convergence of the integral (\ref{deltaint}) as $p_m\rightarrow 1$. Finally we take the limit $n\rightarrow \infty$. From the definitions it is clear that
$(f,\alpha_{-}^nf)\rightarrow (f,\alpha_{-}f)$ and $(f,(\alpha_{-}^n)^{-1}f)
\rightarrow (f,\alpha_{-}^{-1}f)$. These considerations justify the following rigorous definition of the generating functional (\ref{gen1})
\begin{equation}\label{defgen1}
Z(f)/Z(0):=e^{-\frac{1}{2}(f,\alpha_{-}^{-1}f)}\int_{{\cal D'}}F(\phi_++c H_+f)d\mu_{G_+}(\phi_+)/Z(0).
\end{equation}

We now come to a second possible prescription for the AdS/CFT-correspon\-dence. Let us define
\begin{equation}\label{defgen2}
\tilde{Z}(f)/\tilde{Z}(0)=\lim_{z\rightarrow 0}(\tilde{Y}(f)/\tilde{Y}(0))_z:=\lim_{z\rightarrow 0}
\int_{\cal D'}e^{\phi(z^{-\Delta_+}(\delta_z\otimes f))}F(\phi)d\mu_{G_+}(\phi)/
\tilde{Y}(0),
\end{equation}
where $\tilde{Z}(0)=\tilde{Y}(0)$ and $\delta_z\otimes f\in H^{-1}$ is the distribution defined by
$$(\delta_z\otimes f)(g)=\int_{\mathbb{R}^d}f(x)g(z,x)dx, \quad f \in
{\cal S}(\R{d}),\; g\in C_0^\infty(\mathbb{R}_{>0}\times\mathbb{R}^d).
$$
We would like to compare functional (\ref{defgen2}) with the one found in (\ref{defgen1}). To this end we
rewrite (\ref{defgen2}) a little bit using the quasiinvariance of Gaussian measures with respect to shifts by elements from $H^{1}$.
Applying the general result on quasiinvariance, proven e.g. in \cite{BeKo,Bog}, we thus get with $f_z:= z^{-\Delta_+}(\delta_z\otimes f)$
$$
d\mu_{G_+}(\,\cdot -G_+f_z)=e^{\phi((-\Delta+m^2)G_+f_z)}
e^{-\frac{1}{2}(G_+f_z,(-\Delta+m^2)G_+f_z)}d\mu_{G_+}(\,\cdot\,)
$$
$$
=e^{\phi(f_z)}
e^{-\frac{1}{2}(G_+f_z,f_z)}d\mu_{G_+}(\,\cdot\,).
$$
It should be noted that the random field $\phi(f)$ can be extended to all $f\in H^{-1}$.
Using this in (\ref{defgen2}) we arrive at the following expression
\begin{equation}\label{limitb}
(\tilde{Y}(f)/\tilde{Y}(0))_z=
e^{\frac{1}{2}(G_+f_z,f_z)}\int_{\cal D'}F(\phi+G_+f_z)d\mu_{G_+}(\phi)/\tilde{Y}(0).
\end{equation}
Before being able to perform the limit $z\rightarrow 0$ we have to take a closer look at the behavior of the term $(G_+f_z,f_z)$. In appendix A it is shown that in this limit we have to subtract certain divergent terms, more precisely,
$$\int_{\R{d}}\int_{\R{d}}\alpha_+(x,y)f(x)f(y)dxdy=
$$
$$
\lim_{z\rightarrow 0}z^{-d-2\nu}\int_{\R{d}}\int_{\R{d}}G_+(z,x;z,y)f(x)f(y)dxdy -
$$
$$
\frac{1}{(2\pi)^{\frac{d}{2}}}\left(\frac{2^{1-\nu}}{\sqrt{\pi}\Gamma(\nu+\frac{1}{2})}\right)^2\sum_{j=0}^{[\nu]}z^{-2(\nu-j)}(-1)^ja_j\int_{\R{d}}|\hat{f}(k)|^2\bn{k}^{2j}dk.
$$
$$
=:\lim_{z\rightarrow 0}z^{-d-2\nu}\int_{\R{d}}\int_{\R{d}}G_+(z,x;z,y)f(x)f(y)dxdy -({\rm Corr}(z)f,f).
$$
In order to get nontrivial results in the limit we have to regularize the exponential prefactor in (\ref{limitb}) by multiplying it with $
\exp{-(\rm Corr}(z)f,f)$. From (\ref{green}), (\ref{hypergeo}) it is readily seen that, as $z\rightarrow 0$, $G_+f_z$ converges to $H_+f$ uniformly on every bounded subset. Hence assuming
that for some bounded $\Lambda$
\begin{equation}\label{assumption2}
\norm{F(\,\cdot\,+G_+f_z)-F(\,\cdot\,+H_+f)}_{L^1(\mu_{G_+})}\leq {\rm const}\norm{(G_+f_z-H_+f)|_\Lambda}_{L^p},
\end{equation}
for some $p$, we see that the integral in (\ref{limitb}) converges, which shows that the
correct definition for (\ref{defgen2}) reads
$$
\tilde{Z}(f)/\tilde{Z}(0)=\lim_{z \rightarrow 0}e^{-({\rm Corr}(z)f,f)}(\tilde{Y}(f)/\tilde{Y}(0))_z
$$
\begin{equation}\label{limitgen2}
=e^{\frac{1}{2}(\alpha_+f,f)}\int_{{\cal D}'}F(\phi + H_+f)d\mu_{G_+}(\phi)/\tilde{Z}(0).
\end{equation}
In conclusion, we now obtain a proof of the duality conjecture:
\begin{theorem}\label{dualthm}
Suppose that $V$ is such that $F=e^{-V}$ fulfills
(\ref{assumption1}) and (\ref{assumption2}). We then get $Z(f)/Z(0)=
\tilde{Z}(c f)/\tilde{Z}(0)$ when $\nu <\frac{d}{2}$.
\end{theorem}
{\it Proof.} From (\ref{fourierk}) we see that
$\alpha_{-}^{-1}=-c^2\alpha_+$. Compare now (\ref{defgen1}) and
(\ref{limitgen2}) to conclude. $\qquad \Box$\newline Clearly, a
ultra-violet and infra-red regularized local interaction $V_\Lambda$
fulfills the assumptions of the above theorem in any dimension. In
the following section we show that this also holds for the case of
models with polynomial interaction without ultra-violet cut-off on
AdS with $d+1=2$.

\section{$P(\phi)_2$ fields on $\hyp{2}$}
We shall now address the existence of $P(\phi)_2$ models with interaction restricted to some bounded region $\Lambda$. We look at $F_\Lambda(\phi+G_+f_z)=e^{-V_\Lambda(\phi+G_+f_z)}$ with potentials
$$V_\Lambda(\phi)=\int_\Lambda\sum_{j=0}^n :\phi^j(x):_{G_+}f_j(x)dx \equiv \sum_{j=0}^n:\phi^j:(f_j1_\Lambda)=:\left(\sum_{j=0}^n:\phi^j:(f_j)\right)(1_\Lambda),$$ where $:\cdot:_{G_+}$ denotes Wick-ordering with respect to $G_+$.\\
Since $:(\phi+f)^n:_{G_+}(g)=\sum_{j=0}^n\left(n \atop j\right):\phi^j:_{G_+}(gf^{n-j})$, a polynomial interaction $V_\Lambda$ is transformed into such under shifts and we may study the polynomial interaction itself.
\begin{proposition}\label{prointeraction}
Let $V_\Lambda$ be a polynomial interaction as above with $n$ even and let $f_i$ be radial $L^2$-functions. Then
$${\rm (a)} \qquad\|V_\Lambda\|_{L^p(\mu_{G_+})}\leq {\rm const}(p,n)\sum_{i=0}^n\|f_i\|_2,\quad 1\leq p < \infty,$$
and
$${\rm (b)} \qquad \int e^{-V_\Lambda(\phi)}d\mu_{G_+}(\phi)\leq e^{{\rm const}(\|f_n\|_\infty[N(f)+(\ln(M(f)+1))^{\frac{n}{2}}]},$$
where
$N(f)= \sum_{i=0}^{n-1}\|f_i/f_n\|_{n/(n-i)}^{n/(n-i)}$, $M(f)=\sum_{i=1}^n\|f_i\|_{n/(n-i)}.$
$${\rm (c)} \qquad \lim_{z\rightarrow 0}\norm{e^{-V_\Lambda(\,\cdot \, +G_+f_z)}-e^{-V_\Lambda(\,\cdot \, +H_+f_z)}}_{L^p(\mu_{G_+})}=0,\quad 1\leq p<\infty.$$
\end{proposition}
{\it Proof.} The proof is just an adaption of the arguments given in \cite{DiGl} and \cite[Chapter 8]{GlJa} in that Fourier transformation on $\hyp{2}$ is used, see Appendix B. Here we repeat the main steps. Let us consider the expression
\begin{equation}\label{expR}
R_\varepsilon(w,n)=\int \prod_{\mu =1}^N:\phi_\varepsilon(y_\mu)^{n_\mu}:w(y_1,\ldots,y_N)dy, \quad (n=(n_1,\ldots,n_N)\in \mathbb{N}_0^N),
\end{equation}
where $ \phi_\varepsilon(y):=(\phi\star\chi_\varepsilon)(y)$ and
$\chi_\varepsilon(y) := a(\varepsilon)\chi(\varepsilon y),
\,\varepsilon >0,$ is an approximate unity with $\chi \in
C_0^\infty(\hyp{2})$ being a radial function with support in the
unit ball and the factors $a(\varepsilon)$ are chosen such that
$2\pi\int_{\hyp{2}}\chi_\varepsilon(r)\sinh rdr=1$ for all
$\varepsilon$. We assume that the support of $w$ is contained in
$B_1\times \cdots \times B_N$, where the $B_i$ are balls in
$\hyp{2}$. The integral of (\ref{expR}) with respect to $d\mu_{G_+}$
can be calculated as a sum of vacuum graphs. The graphs in the
present case are built as follows: Consider $N$ vertices each having
$n_\mu$ $(1\leq \mu \leq N)$ legs and combine arbitrary pairs of
legs from different vertices to lines to obtain a graph. Vacuum
graphs comprise the subset of graphs where all legs are paired.
Denoting $[I]$ the set of all legs and $\Gamma_0(I)$ the set of all
vacuum graphs, the integral of (\ref{expR}) can be estimated as
$(0\leq \rho \leq 1)$
$$
\left|\int R(w,n)_\varepsilon d\mu_{G_+}\right|\leq M(\rho,n,G_+)\|w\|_2
$$
\begin{equation}
\times\label{estR}\prod_{(\mu,k)\in[I]}\norm{\chi_{\mu,k}}_1^{1-\rho}\norm{\hat{\chi}_{\mu,k}(m^2+{\T \frac{1}{4}}+\lambda^2)^{-\delta/4}}^\rho_\infty,
\end{equation}
where $\chi_{\mu,k}=\chi_\varepsilon$ and the constant is given as
$(n_\ast=\sup_\mu n_\mu,\, p'=\frac{p}{p-1},\,\delta\leq 2)$
\begin{equation}\label{constantM}
M(\rho,n,G_+)=|B|\sum_{G\in\Gamma_0(I)}\prod_{l\in G}\norm{(\zeta_{l_-}\otimes\zeta_{l_+})G_+}^{1-\rho}_{2n_\ast}\norm{(\zeta_{l_-}\otimes\zeta_{l_+})G_+}^{\rho}_{{\cal B}_{(2n_\ast)',\delta}}.
\end{equation}
The tuples $(l_-,l_+)$ refer to some ordering of vertices (smaller, larger) and $\zeta_{\mu,k}=\zeta_\mu$ is any radial $C_0^\infty(\hyp{2})$ function which is identically one on $\{x:{\rm dist}(x,B_\mu)\leq 1\}$.
In (\ref{constantM}) we have used the norm
$$\norm{\zeta \psi}_{{\cal B}_{r,\delta}}:=\norm{(1+|\lambda|^2)^{\frac{\delta}{2}}(\widehat{\zeta\psi})}_{L^r}.$$
Using the coordinate characterization of Sobolev spaces we see that $G_+\in H^{-1}\times H^{-1}$ implies $\zeta G_+\in H^{-1}\times H^{-1}.$ The Fourier space characterization of Sobolev spaces then shows that the norms $\norm{(\zeta_{l_-}\otimes\zeta_{l_+})G_+}^{\rho}_{{\cal B}_{(2n_\ast)',\delta}}$ are finite.
Moreover, using (\ref{estR}) and noting that
$\norm{\hat{\chi}_\varepsilon(m^2+{\T \frac{1}{4}}+\lambda^2)^{-\delta/4}}_\infty\leq {\rm const}\norm{\hat{\chi}_\varepsilon}_\infty\leq {\rm const}\norm{\chi_\varepsilon}_1
$ and
$\norm{(\hat{\chi}_\varepsilon-\hat{\chi}_{\varepsilon'})(m^2+{\T\frac{1}{4}}+\lambda^2)^{-\delta/4}}_\infty \leq O(1)(\varepsilon\wedge \varepsilon')^{-\delta/2}$, see Appendix B, one derives
\begin{equation}\label{estR1}
\norm{R_\varepsilon(w,n)}_{L^p(\mu_{G_+})}\leq {\rm const}(p,n)\norm{w}_2
\end{equation} and
\begin{equation}\label{estdiff}
\norm{R_\varepsilon(w,n)-R_{\varepsilon'}(w,n)}_{L^p(\mu_{G_+})}\leq {\rm const}(p,n)(\varepsilon\wedge \varepsilon')^{-\delta/2}\norm{w}_2.
\end{equation}
The latter inequalities show that $R_\varepsilon$ is a
Cauchy-sequence in $L^p(\mu_{G_+})$ with limit $R(w,n)$ obeying the
bound (\ref{estR1}). Applying this to the special case
$R(w,n)=V_\Lambda$ we get statement $(a)$. In 2 dimensions there is
just a logarithmic singularity $G_+(\ul{x},\ul{x}')\sim {\rm
const}|\ln d(\ul{x},\ul{x}')|$ for small distances. With the aid of
(\ref{estR1}) and (\ref{estdiff}), by employing the arguments given
in \cite[Theorem 8.6.2]{GlJa}, we see that also (b) holds true. In
order to prove (c), we write
$$e^{-V(\,\cdot \, +G_+f_z)(1_\Lambda)}-e^{-V(\,\cdot \, +H_+f)(1_\Lambda))}$$
$$=\int_0^1 e^{-V(\,\cdot \,+G_+f_z)(s1_\Lambda)}\times
$$
$$
(V(\,\cdot\, +H_+f)(1_\Lambda)-V(\,\cdot\, +G_+f_z)(1_\Lambda))e^{-V(\,\cdot\, +H_+f)((1-s)(1_\Lambda))}ds.$$
The $L^p(\mu_{G_+})$-norm of the latter integral can be estimated as
$$\sup_{0\leq s\leq 1}\left(\norm{e^{-V(\,\cdot \,+G_+f_z)(s1_\Lambda)}}_{L^{3p}(\mu_{G_+})}\norm{e^{-V(\,\cdot\, +H_+f)((1-s)(1_\Lambda))}}_{L^{3p}(\mu_{G_+})}\right)
$$
$$\times\norm{V(\,\cdot\, +G_+f_z)(1_\Lambda)-V(\,\cdot\, +H_+f)(1_\Lambda)}_{L^{3p}(\mu_{G_+})},
$$
which by $(a)$ and $(b)$ proves the assertion.
$\qquad \Box$ \newline
\section{Reflection positivity and invariance}
In this section we probe the functional integrals for reflection
positivity and conformal invariance. These two properties are
essential to qualify them as providing us a conformal field theory
on the boundary. The following considerations are valid for $d\geq
1$, if we assume that a local (hence reflection positivity
preserving) interaction exists for bounded $\Lambda$ and limits of
the generating functionals exist for $\Lambda \nearrow {\mathbb
H}^{d+1}$. The existence and related questions of uniqueness are
left to future work.

For simplicity let us consider the reflection $\theta$ with respect to coordinate $x_1$ of $\R{d}$, i.e. $\theta(x_1,x_2,\ldots,x_d)=(-x_1,x_2,\ldots,x_d)$. Let $\Lambda\subset\R{d+1}$ be reflection-symmetric, where the action of $\theta$ is extended to
$\R{d+1}$ via $\theta(\underline{x})=(z,\theta{x})$. We want to verify that the integrals $(\tilde{Y}(f)/\tilde{Y}(0))_z$ are reflection positive, i.e. the finite matrix $m_{ij}=(\tilde{Y}(f_i+\theta f_j)/\tilde{Y}(0))_z$ is positive-semidefinite for arbitrary $f_i\in {\cal S}(\R{d})$ with support at $x_1>0$. Note that our formulation of reflection positivity refers to the Laplace transform of measures and not to their Fourier transform. For local interactions the latter are restrictions of a reflection positive generating functional, see \cite{GlJa,Di}, in the sense that
$$(\tilde{Y}(f)/\tilde{Y}(0))_z=\lim_{n\rightarrow \infty}\int_{\cal D'}e^{\phi(z^{-\Delta_+}g_n)}F_\Lambda(\phi)d\mu_{G_+}(\phi)/
\tilde{Y}(0),$$
for a sequence $g_n$ converging to $\delta_z\otimes f$ in $H^{-1}$.
In order that $\tilde{Z}(f)/\tilde{Z}(0)$ be reflection positive, we need that the correcting factor $\exp(-{\rm Corr}(z)f,f)$ is reflection positive. But the covariances in ${\rm Corr}(z)$ are given by the inverse Fourier transforms of $\bn{k}^{2j}$ which equal ${\rm const}(j)(-1)^{2j}\delta^{(2j)}(|x|)$, see \cite{GeSh}. According to \cite[Theorem 6.2.2]{GlJa} the generating functional of a Gaussian measure $d\mu_C$ is reflection positive if the covariance satisfies $(\theta f,Cf)\geq 0$ for all $f$ supported at positive $x_1$. Using an approximation argument it is sufficient to check this property for functions of the form $f=f_rf_\varphi$, with $f_r\in C_0^\infty(\mathbb{R}_{> 0})$ and $f_\varphi \in C^\infty(\mathbb{S}^{d-1}_+)$, where $\mathbb{S}^{d-1}_+=\{x\in \R{d}:|x|=1,\,x_1>0\}$. In this case $(\theta (f_rf_\varphi),\delta^{(2j)}(f_rf_\varphi))=((\theta f_r)^{(j)}\theta f_\varphi,f_r^{(j)}f_\varphi)=0$, hence the claim. It follows that reflection positivity holds also for $z\rightarrow 0$ and then for $\Lambda\nearrow \mathbb{H}^{d+1}$.

The basic implication of the AdS/CFT correspondence is that covariance of the bulk functional integral translates into a conformal invariance on the boundary. On geometrical grounds the isometry group ${\rm Iso}(\hyp{d+1})$ acts by conformal transformations on the boundary, see \cite{Kni}. Here we allow also non-orientation preserving isometries and conformal transformations. This means in particular that
\begin{equation}\label{baction}
g^\ast dx={\rm det}\left(\frac{\partial g(x)}{\partial x}\right)dx,
\end{equation}
where $dx$ is the standard volume form on $\R{d}$ and $\partial g(x)/\partial
x$ denotes the Jacobian matrix. In order to take into consideration
the transformations (\ref{baction}), we regard our functionals as
functions of $d$-forms $\omega$ with compact support, i.e.,
$\omega=fdx$ with $f\in C_0^\infty(\R{d})$ such that the support of $f$ doesn't contain a point, which potentially is mapped to infinity by $g$.

Suppose that
$\tilde{Z}_{\lim}(\omega):=\lim_{\Lambda \nearrow {\mathbb H}^{d+1}}
\tilde{Z}(f)/\tilde{Z}(0)$ exists uniquely, then conformal
invariance means the property that
\begin{equation}\label{cinv}
\tilde{Z}_{\lim}(g\omega)=\tilde{Z}_{\lim}(\lambda_g\cdot \omega),\quad g\in {\rm Iso}(\hyp{d+1}),
\end{equation}
with action $g\omega:=g^{{-1}\ast}\omega$ and scale factor $\lambda_g(x)=\left|{\rm det}\left(\frac{\partial g(x)}{\partial x}\right)\right|^{-\frac{\Delta_+}{d}}$.
For (\ref{cinv}) to hold the bulk-to-boundary propagator has to fulfill the following intertwining property.
\begin{lemma}
For $g\in {\rm Iso}(\mathbb{H}^{d+1})$ let $g(z,x)=(z_g(z,x),x_g(z,x))\equiv (z_g,x_g)$ denote the action of $g$. Then with $g(x)=\lim_{z\rightarrow 0}x_g(z,x)$ we have
$$
H_+(g(z,x);x')=\left|{\rm det}\left(\frac{\partial g^{-1}(x')}{\partial x'}\right)\right|^{\frac{\Delta_+}{d}}H_+(z,x;g^{-1}(x'))
$$
\end{lemma}
{\it Proof.} We note that $H_+(z,x;x')=\lim_{z'\rightarrow 0}z'^{-\Delta_+}\left(\frac{zz'}{(z-z')^2+(x-x')^2}\right)^{\Delta_+}
$
Now, \\
$\frac{zz'}{(z-z')^2+(x-x')^2}$ is invariant with respect to isometries and therefore
$$H_+(g(z,x);x')=\lim_{z' \rightarrow 0}z'^{-\Delta_+}
\left(\frac{zz'_{g^{-1}}}{(z-z'_{g^{-1}})^2+(x-x'_{g^{-1}})^2}\right)^{\Delta_+}$$
$$=\lim_{z' \rightarrow 0}(z'_{g^{-1}})^{-\Delta_+}\left(\frac{z'}{z'_{g^{-1}}}\right)^{-\Delta_+}\left(\frac{zz'_{g^ {-1}}}{(z-z'_{g^{-1}})^2+(x-x'_{g^{-1}})^2}\right)^{\Delta_+}.
$$
In order to see the effect of the transformation $g^{-1}$ we use its action on the isometric model of $\hyp{d+1}$ given by
$$
\mathbb{L}^{d+1}:=\{\zeta\in \mathbb{M}^{d+1,1}\,|\,\zeta_1^2+\cdots+\zeta_{d+1}^2-\zeta_{d+2}^2 =-1,\, \zeta_{d+2}>0\},
$$
equipped with the metric induced from Minkowski space $\mathbb{M}^{d+1,1}$ with metric $d\zeta_1^2+\cdots+d\zeta_{d+1}^2-d\zeta_{d+2}^2.$ For $\mathbb{L}^{d+1}$ the isometry group is by definition $O^+(d+1,1)$ and the isometry map $\eta:\hyp{d+1}\rightarrow \mathbb{L}^{d+1}$ is given by
$$\zeta_i= \frac{x_i}{z},\quad 1\leq i \leq d,
$$
$$
\zeta_{d+1}=-\frac{1}{2z}(z^2+x^2-1),\quad\zeta_{d+2}=\frac{1}{2z}(z^2+x^2+1).
$$

with inverse
$$
z=\frac{1}{\zeta_{d+1}+\zeta_{d+2}},\quad x_i=\frac{\zeta_i}{\zeta_{d+1}+\zeta_{d+2}}.
$$
Thus, an arbitrary isometry on $\hyp{d+1}$ can be cast into the form
$$\eta^{-1} \circ g \circ \eta,\quad g\in O^+(d+1,1).$$
Using this fact, one easily shows that $z'_{ g^{-1}},\,\partial z'_{g^{-1}}/\partial x_i'$ and $\partial x'_{i{g^{-1}}}/\partial z'$ tend to zero as $z'\rightarrow 0$, whereas $\partial x'_{g^{-1}}/\partial x'\rightarrow \partial g^{-1}(x')/\partial x'$ and $\partial z'_{g^{-1}}/\partial z'\sim z'_{g^{-1}}/z'$. Moreover, invariance of the volume measure $z^{-d-1}dzdx$, up to a possible sign, implies
$$z'^{-d-1}=(z'_{g^{-1}})^{-d-1}\left|{\rm det}\left(\frac{\partial g^{-1}(z',x')}{\partial (z',x')}\right)\right|.$$
Combining all this, gives
$$
\lim_{z'\rightarrow 0}\left(\frac{z'_{g^{-1}}}{z'}\right)=\left|{\rm det}\left(\frac{\partial g^{-1}(x')}{\partial x'}\right)\right|^{\frac{1}{d}},
$$ which shows the statement of the Lemma. $\quad \Box$ \\
We may summarize the above findings in
\begin{theorem} Let $\theta$ be the reflection with respect to a hyperplane of $\R{d}$ containing $0$.
If the limit $\tilde{Z}_{\lim}(\omega):=\lim_{\Lambda \nearrow {\mathbb H}^{d+1}} \tilde{Z}(f)/\tilde{Z}(0)$ exists for a sequence of reflection-invariant $\Lambda's$, then it is reflection positive, in the sense that the finite matrix $M_{ij}=\tilde{Z}_{\lim}( \omega_i+\theta\omega_j),\, \theta \omega_i=(\theta f_i)dx,$ is positive-semidefinite for arbitrary $\omega_i$ with $f_i\in {\cal S}(\R{d})$, having support in the positive half-space. Moreover, if the limit exists uniquely, then conformal invariance holds in the sense of equation (\ref{cinv}).
\end{theorem}
{\it Remark 3.} When the functional $\tilde{Z}_{\lim}$ is analytic
at $0$, reflection positivity of $\tilde{Z}_{\lim}$ entails
reflection positivity of the corresponding Schwinger functions
$(S_n)_{n\in \mathbb{N}_0}$. This also holds in the case when
$\tilde Z_{\rm lim}$ is not stochastically positive, see \cite[Prop.
6.1]{Go}. Note that the correction term $({\rm Corr}(z)f,f)$
potentially destroys stochastic positivity in the limit $z\to 0$.\\
{\it Remark 4.} The case of conformal symmetry is not treated in the
standard version of the Osterwalder-Schrader reconstruction theorem,
cf. e.g. \cite{OsSch1,OsSch2}. The required extension of the reconstruction
theorem can easily be accomplished in the same way as the
relation of rotation- and Lorentz invariance, writing for example,
dilatation invariance infinitesimally as
$\left(\sum_{j=1}^nx_j\cdot\nabla_{x_j}-n\Delta\right)S_n(x_1,\ldots,x_n)=0$,
$\Delta$ being the conformal weight, and then representing $S_n$ as
a Laplace transform of the Fourier transformed Wightman function.
Via taking the differential operator under the integral transform
and integration by parts, dilatation invariance of the Fourier
transformed Wightman functions with weight $\Delta-d$ follows which
is equivalent to dilatation invariance of Wightman functions with
weight $\Delta$. The same argument applies for special conformal transformations.

We have thus completed the proof of AdS/CFT for Euclidean quantum
fields up to the infra-red problem $\Lambda\nearrow\mathbb{H}^{d+1}$
(for $d+1=2$). Due to the different nature of source terms, which
include bulk-to-boundary propagators that increase if one approaches
the conformal boundary in the direction of the source, this
infra-red problem is different from, and probably much harder as,
the related one in \cite{GlJa} where sources are rapidly decaying.
We will come back to this point elsewhere.
\begin{appendix}
\section{Divergencies in $\lim_{z\rightarrow 0}z^{-2\Delta_+}
(G_+f_z,f_z)$}
In investigating this limit we shall use the following integral representation, see
\cite{BBMS,LiTs},
$$
G_+(z,x;z',y)=(zz')^{d/2}\frac{1}{(2\pi)^{\frac{d}{2}}}\int_0^\infty\int_{\mathbb{R}^d}\frac{1}{\omega^2+\bn{k}^2}e^{ik(x-y)}dkJ_\nu(z\omega)J_\nu(z'\omega)\omega d\omega
$$
\begin{equation}\label{greenex}
=(zz')^{d/2}\frac{1}{(2\pi)^{\frac{d}{2}}}\int_0^\infty
C_\omega(x-y)J_\nu(z\omega)J_\nu(z'\omega)\omega d\omega,
\end{equation}
where $C_\omega$ is the integral kernel of $(-\Delta+\omega^2)^{-1}$ in $\R{d}$.
In addition, for $\rm{Re}\nu >-\frac{1}{2}$, $J_\nu$ can be represented as
$$J_\nu(u)=\frac{2^{1-\nu}}{\sqrt{\pi}\Gamma(\nu+\frac{1}{2})}u^\nu\int_0^1(1-t^2)^{\nu-\frac{1}{2}}\cos(ut)dt.
$$
Then with $f\in {\cal S}(\R{d})$ we get
$$\int_{\R{d}}\int_{\R{d}} G_+(z,x;z,y)f(x)f(y)dxdy =
\frac{z^{d+2\nu}}{(2\pi)^{\frac{d}{2}}}\left(\frac{2^{1-\nu}}{\sqrt{\pi}\Gamma(\nu+\frac{1}{2})}\right)^2\times
$$
$$
\int_0^\infty\int_{\R{d}} \frac{|\hat{f}(k)|^2}{\omega^2+\bn{k}^2}dk\left(\int_0^1(1-t^2)^{\nu-\frac{1}{2}}\cos(z\omega t)dt\right)^2\omega^{2\nu+1}d\omega.
$$
Employing the geometric series expansion
$$\frac{1}{\omega^2+\bn{k}^2}=\frac{1}{\omega^2}\left(\frac{1}{1+\bn{k}^2/\omega^2}\right)=\sum_{j=0}^{[\nu]}(-1)^j\frac{\bn{k}^{2j}}{\omega^{2j+2}}+\frac{(-\bn{k}^2/\omega^2)^{[\nu]+1}}{\omega^2+\bn{k}^2}
$$
we obtain
$$z^{-d-2\nu}\int_{\R{d}}\int_{\R{d}} G_+(z,x;z,y)f(x)f(y)dxdy =\frac{1}{(2\pi)^{\frac{d}{2}}}\Bigg(\frac{2^{1-\nu}}{\sqrt{\pi}\Gamma(\nu+\frac{1}{2})}\Bigg)^2\times$$
$$ \Bigg\{\sum_{j=0}^{[\nu]}(-1)^j\int_0^\infty \omega^{2(\nu-j)-1}\left(\int_0^1\cos(z\omega t)(1-t^2)^{\nu -\frac{1}{2}}dt\right)^2d\omega\int_{\R{d}}|\hat{f}(k)|^2\bn{k}^{2j}dk
$$
$$
+(-1)^{[\nu]+1}\int_0^\infty\int_{\R{d}}\frac{|\hat{f}(k)|^2}{\omega^2+\bn{k}^2}\bn{k}^{2[\nu]+2}dk\;\times
$$
\begin{equation}\label{secterm}
\omega^{2(\nu-[\nu])-1}\left(\int_0^1\cos(z\omega t)(1-t^2)^{\nu-\frac{1}{2}}dt\right)^2d\omega\Bigg\}.
\end{equation}
On the one hand, the terms
$$\int_0^\infty \omega^{2(\nu-j)-1}\left(\int_0^1\cos(z\omega t)(1-t^2)^{\nu -\frac{1}{2}}dt\right)^2d\omega$$
$$=z^{-2(\nu-j)}\int_0^\infty \left(\int_0^1\cos(\omega t)(1-t^2)^{\nu-\frac{1}{2}}dt\right)^2\omega^{2(\nu-j)-1}d\omega=:z^{-2(\nu-j)}a_j$$
diverge as $z\rightarrow 0$. On the other hand, using dominated convergence, one can show that the last term
in (\ref{secterm}), for $z\rightarrow 0$ converges to constant times
$$\left(\int_0^1(1-t^2)^{\nu -\frac{1}{2}}dt\right)^2\int_0^\infty\int_{\R{d}}\frac{|\hat{f}(k)|^2}{\bn{k}^2+\omega^2}\bn{k}^{2[\nu]+2}dk\,\omega^{2(\nu-[\nu])-1}d\omega.
$$
Using the formula
$$\int_0^1 t^{2a+1}(1-t^2)^b dt=\frac{1}{2}\left(\frac{\Gamma(a+1)\Gamma(b+1)}{\Gamma(a+b+2)}\right)
$$ with $a=-\frac{1}{2}, b=\nu-\frac{1}{2}$
we thus get
$$\lim_{z\rightarrow 0}\Bigg\{z^{-d-2\nu}\int_{\R{d}}\int_{\R{d}}G_+(z,x;z,y)f(x)f(y)dxdy-
$$
$$\frac{1}{(2\pi)^{\frac{d}{2}}}\left(\frac{2^{1-\nu}}{\sqrt{\pi}\Gamma(\nu+\frac{1}{2})}\right)^2\sum_{j=0}^{[\nu]}z^{-2(\nu-j)}(-1)^ja_j\int_{\R{d}}|\hat{f}(k)|^2\bn{k}^{2j}dk\Bigg\}=$$
\begin{equation}\label{reglimit}
\frac{1}{(2\pi)^{\frac{d}{2}}}\left(\frac{2^{-\nu}\Gamma(\frac{1}{2})}{\sqrt{\pi}\Gamma(\nu+1)}\right)^2(-1)^{[\nu]+1}\int_0^\infty\int_{\R{d}}\frac{|\hat{f}(k)|^2}{\omega^2+\bn{k}^2}\bn{k}^{2[\nu]+2}dk\,\omega^{2(\nu-[\nu])-1}d\omega.
\end{equation}
Let us perform the $\omega$-integration in (\ref{reglimit}) first. With the aid of
$$\int_0^\infty \frac{x^{a-1}}{1+x^b}dx=\frac{\pi}{b\sin(a\pi/b)},\quad 0<a<b,$$
where $a=2(\nu-[\nu]),b=2,$ we get for the integral
$$
\int_{\R{d}}\frac{\pi \bn{k}^{2(\nu-[\nu])-2}}{2\sin\left(\frac{2(\nu-[\nu])\pi}{2}\right)}|\hat{f}(k)|^2\bn{k}^{2[\nu]+2}dk=(-1)^{[\nu]}\frac{\pi}{2\sin\nu\pi}\int_{\R{d}}|\hat{f}(k)|\bn{k}^{2\nu}dk,
$$
and therefore (\ref{reglimit}) simplifies to
\begin{equation}\label{endalpha}
-\frac{1}{(2\pi)^{\frac{d}{2}}}\frac{\pi}{2\sin(\nu\pi)}\left(\frac{1}{2^\nu\Gamma(\nu+1)}\right)^2\int_{\R{d}}|\hat{f}(k)|^2\bn{k}^{2\nu}dk.
\end{equation}
Comparing (\ref{endalpha}) with (\ref{fourierk}) and exploiting relations $\Gamma(\nu)\Gamma(1-\nu)=\pi/\sin(\nu\pi)$ and $\Gamma(1-\nu)=-\nu\Gamma(-\nu)$ we see that the latter expression equals $(f,\alpha_+f)$.
\section{Fourier and spherical Fourier transform on $\hyp{d}$}
Hyperbolic spaces belong to the class of Riemannian symmetric spaces which can be represented in the form $X=G/K$ with $G$ a noncompact semisimple Lie group and $K$ a maximal compact subgroup, i.e. $\hyp{d+1}\simeq SO_0(d+1,1)/SO_0(d+1)$. For these type of spaces there is an analogue of the Fourier transform in $\mathbb{R}^d$. Let $\mathfrak{g}=\mathfrak{k}\oplus\mathfrak{p}$ be the Cartan decomposition of the Lie algebra $\mathfrak{g}$ of $G$. Then we have the following Iwasawa decomposition $\mathfrak{g}=\mathfrak{k}\oplus\mathfrak{a}\oplus \mathfrak{n}$, where $\mathfrak{a}$ is a maximal abelian subspace of $\mathfrak{p}$, $\mathfrak{n}:=\bigoplus_{\alpha \in \Sigma_+}\mathfrak{g}_\alpha$ with $\Sigma_+$ being a choice of positive roots with respect to $(\mathfrak{g},\mathfrak{a})$. The norm induced from the Killing-form on $\mathfrak{p}$ will be denoted by $\norm{\cdot}$. There is a corresponding Iwasawa decomposition for the Lie group $G=KAN=NAK$ and every $g\in G$ can be written as $g=k(g)\exp H(g)n(g)$ with unique elements $k(g) \in K, H(g) \in \mathfrak{a}, n(g) \in N.$ Let $M$ denote the centralizer of $A$ in $K$, $B:=K/M$ and let $A(x,b)\in \mathfrak{a}$ be the vector $A(x,b):=A(k^{-1}g)$ for $x=gK\in X$ and $b=kM\in B$. The Fourier transform of a function $f \in C_0^\infty(X)$ is now defined as \cite{He}
\begin{equation}\label{fourier}
\hat{f}(\lambda,b):=\int_X f(x)e^{(-i\lambda +\rho)A(x,b)}dx,\quad \lambda \in \mathfrak{a}^\ast_\mathbb{C},\, b\in B,
\end{equation}
where $\rho =\frac{1}{2}\sum_{\alpha\in \Sigma_+}m_\alpha\alpha,\,m_\alpha={\rm dim}\mathfrak{g}_\alpha$.
Let us have a closer look at the space $\hyp{2}$ which can be represented as the open disk
$D:=\{w \in \mathbb{C}:|w|<1\}$,
equipped with the Riemannian metric $g_D=4(1-|w|^2)^{-2}(dw_1^2+ dw_2^2)$, which in turn is diffeomorphic to the homogenous space $G/K$ where the Lie group
$$G=SU(1,1)=\left\{g=\left(
\begin{array}{cc}
a & \;b \\
\bar{b} &\; \bar{a}
\end{array}
\right):|a|^2-|b|^2=1\right\}
$$
acts on $D$ by
$$g\cdot w=\frac{aw+b}{\bar{b}w+\bar{a}}$$
and the isotropy group of $0$ is $K=SO(2).$
In this picture the Fourier transform of a function on $D$ is given by
$$
\hat{f}(\lambda,b)=\int_D f(w)e^{(-i\lambda+\frac{1}{2})\langle w,b\rangle}d\sigma(w),\qquad \lambda \in \mathbb{C},\, b \in \partial D=B,
$$
where $d\sigma$ is the volume form related to $g_D$ and $\langle w,b \rangle$ denotes the geodesic distance from $0$ to the circle which passes through $w$, and at $b$, is tangential to the boundary $\partial D$ of $D$.
The spherical Fourier transform is defined by
\begin{equation}\label{spherfour}
\hat{f}(\lambda)=\int_D f(w)\phi_{-\lambda}(w)d\sigma(w),
\end{equation}
where $\phi_\lambda$ is the spherical function
$$\phi_\lambda(w)=\int_{\partial D}e^{(i\lambda+\frac{1}{2})\langle w,b\rangle}db.
$$
In the general case spherical functions are given by $\phi_\lambda(g)=\int_K e^{(i\lambda +\rho)A(k^{-1}g)}dk$ and obey $\phi_\lambda(e)=1$ and $-\Delta\phi_\lambda=(\norm{\lambda^2}+\norm{\rho}^2)\phi_\lambda$ . We notice that for radial functions $f$, i.e. $f(w)=f(|w|)$, the transform (\ref{spherfour}) may be written as
$$
\hat{f}(\lambda)=2\pi\int_0^\infty f(\tanh {\T \frac{r}{2}})\phi_\lambda(\tanh \T
{\frac{r}{2}})\sinh rdr,\qquad (r=d(0,w)),
$$
Moreover, since $e^{\langle w,b\rangle}=\frac{1-|w|^2}{|w-b|^2}$, with the substitutions $w=\tanh {\T\frac{r}{2}},\, b=e^{i\theta}$, we may write
$$
\phi_\lambda(\tanh {\T \frac{r}{2}})=\frac{1}{2\pi}\int_{-\pi}^\pi(\cosh r-\sinh r \cos \theta)^{-(i\lambda +\frac{1}{2})}d\theta,
$$ and
setting further $u=\tanh \T{\frac{1}{2}}\theta,\,\frac{1}{2}d\theta=(1+u^2)^{-1}du$, we get
\begin{equation}\label{spherpol}
\phi_\lambda(\tanh {\T \frac{r}{2}})=\frac{1}{\pi} \int_{-\infty}^\infty \left(\cosh r +\sinh r\frac{1-u^2}{1+u^2}\right)^{(i\lambda -\frac{1}{2})}\frac{du}{1+u^2}.
\end{equation}
Because of the group structure we may consider the convolution
$$(f_1\star f_2)(g\cdot o):=\int_G f_1(h\cdot o)f_2(h^{-1}g\cdot o)dh,\quad o=eK.$$
For radial functions $f_1, f_2$ one gets
$$(\widehat{f_1\star f_2})(\lambda)=\hat{f}_1(\lambda)\hat{f}_2(\lambda),$$
whenever both sides exist.
We also need the following estimate
$$\norm{(\hat{\chi}_\varepsilon-\hat{\chi}_{\varepsilon'})(\lambda^2+{\T \frac{1}{4}}+m^2)^{-\frac{\delta}{4}}}_\infty \leq O(1)(\varepsilon\wedge\varepsilon')^{-\frac{\delta}{2}}.$$
To see this let us regard $\hat{\chi}_\varepsilon$ as a function of $\frac{\lambda}{\varepsilon}$ by setting $g_\varepsilon(\lambda/\varepsilon) :=\hat{\chi}_\varepsilon(\lambda)$. Then $\frac{d}{d(\lambda/\varepsilon)} (g_\varepsilon)=\varepsilon \frac{d}{d\lambda}(\hat{\chi}_\varepsilon)$ and with the aid of (\ref{spherpol}) and the substitution $y=\varepsilon r$ we get that
$$|\hat{\chi}_\varepsilon(\lambda) -\hat{\chi}_{\varepsilon'}(\lambda)|\leq O(1)\left(\frac{|\lambda|}{\varepsilon\wedge\varepsilon'}\right)^{\frac{\delta}{2}}$$
 for $|\lambda| \leq \varepsilon\wedge\varepsilon'$ and
$$|\hat{\chi}_\varepsilon(\lambda) -\hat{\chi}_{\varepsilon'}(\lambda)|\leq O(1)$$
 for $|\lambda| > \varepsilon\wedge\varepsilon'$.
\section{Sobolev spaces}\label{sobolev}
In this section we introduce Sobolev spaces on hyperbolic spaces. For $\beta \geq 0$ let us define the Sobolev space of order $\beta$ as in \cite{Str}
$$H^\beta :=\left\{u \in L^2(\hyp{d+1}): u =(-\Delta+m^2)^{-\frac{\beta}{2}}v,
\;v \in L^2(\hyp{d+1})\right\}$$
with $\|u\|_{H^\beta}:=
\|v\|_{L^2(\hyp{d+1})}. $ For $\beta < 0$ we define
$$H^\beta
:= \left\{u \in {\cal D}': u = (-\Delta+m^2)^k v, v \in H^{2k+\beta}\; \mbox{with}\; k \;\mbox{such that}\;  2k + \beta >0\right\},$$
and norm $\|u\|_{H^\beta}:=\|v\|_{H^{2k+\beta}}$. By definition the maps $-\Delta+m^2 :H^\beta \rightarrow H^{\beta -2}$ and $(-\Delta+m^2)^{-1}:H^\beta \rightarrow H^{\beta +2}$ are isomorphisms of Hilbert spaces. The spaces $H^\beta$ can be identified with the completion of $C_0^\infty(\hyp{d+1})$ in the norm $\norm{f}_{H^\beta}=\norm{(-\Delta+m^2)^{\frac{\beta}{2}}f}_{L^2(\hyp{d+1})}.$

In section 3 we have used the distribution $f_z=\delta_z \otimes f$ with $f \in {\cal S}(\mathbb{R}^d).$ Using the explicit expression (\ref{greenex}) and a proper
smoothing with an approximate unit one sees that $f_z\in H^{-1}.$

A second equivalent definition of Sobolev spaces uses local
coordinates, see \cite{tar}. For this one first considers the space
$H^\beta_o$ of distributions which are supported in a ball of fixed
radius $r,$ $B(o,r),$ around some fixed point $o$ equipped with
geodesic coordinates and defines the norm $\|u \|_{H_o^\beta}$ as
the pull-back of the $H^\beta(\mathbb{R}^{d+1})$ norm in the chosen
coordinates. For another $a \in \hyp{d+1}$ and distribution $f$
supported in $B(a,r)$ one defines $\|f\|_{H^\beta_a}:=\|f\circ
g\|_{H_o^\beta}$ where $g$ is an isometry with $g(o)=a$. Then points
$(a_k)_{k\in \mathbb{N}}$ are chosen in order to obtain a locally
finite covering by the balls $B(a_k,r)$. Finally, employing a
partition of unity $(\varphi_k)_{k\in \mathbb{N}}$ w.r.t. the balls
$B(a_k,r)$ one says that $u\in H^\beta(\hyp{d+1})$ if
$$\sum_k \|\varphi_k u \|_{H^\beta_{a_k}}< \infty.$$

A third definition can be given using Fourier transforms, where we follow \cite{EgOk}. For this we define the Schwartz space ${\cal S}(\hyp{d+1})={\cal S}(X)$, consisting of complex-valued $C^\infty$-functions $f$ on $X$ satisfying
$$\tau_{D,m}(f)=\sup_{g\in G}(1+|g|)^m\phi_0(g)^{-1}|Df(g)|<\infty,$$ for all $m \in \mathbb{N}_0$ and differential operators $D$ invariant under the left action of $G$. The norm of $g$ is defined as
$|g|=|\exp Xk|=\|X\|$, $X\in \mathfrak{p},k\in K$. The space ${\cal S}(X)$ becomes a Fr\'{e}chet space when topologized by means of the seminorms $\tau_{D,m}.$ Let ${\cal S}(\mathfrak{a}^\ast\times K/M)$ be the complex-valued $C^\infty$-functions on $\mathfrak{a}^\ast \times K/M$ such that
$$\nu_{E,J,r}(f)=\sup_{\lambda,kM}(1+\|\lambda\|)^r|(EJf)(\lambda,kM)|<\infty,$$
for all differential operators $E$ on $\mathfrak{a}^\ast$ and $J$ invariant on
$K/M$ and $r\in \mathbb{N}_0$. With these seminorms ${\cal S}(\mathfrak{a}^\ast\times K/M)$ becomes a Fr\'{e}chet space. The Fourier transform (\ref{fourier}) establishes a topological isomorphism between ${\cal S}(X)$ and ${\cal S}(\mathfrak{a}^\ast\times B)_W={\cal S}(\mathfrak{a}^\ast_+\times B)$, where the subscript $W$ denotes the quotient space under the action of the Weyl group on $\mathfrak{a}^\ast$. Moreover, the Fourier transform extends to an isometry of $L^2(X)$ onto $L^2(\mathfrak{a}^\ast_+\times B,|c(\lambda)|^{-2}d\lambda db)$, where $c(\lambda)$ is the Harish-Chandra $c$-function. From the property $(-\widehat{\Delta f})(\lambda,b)=(\norm{\lambda}^2+\norm{\rho}^2)\hat{f}(\lambda,b)$ we see that $H^\beta(X)$ equals the space of $u\in {\cal S}(X)'$ such that
$$\int_{\mathfrak{a}^\ast_+}\int_B |\hat{u}|^2(\norm{\lambda}^2+\norm{\rho}^2+m^2)^\beta|c(\lambda)|^{-2} d\lambda db<\infty.$$
It should be noted that in the case $\hyp{2}$ we have $\mathfrak{a}^\ast_+=\mathbb{R}_+,\, \norm{\rho}^2=\frac{1}{4}$ and \\$|c(\lambda)|^{-2}=(2\pi)^{-1}\lambda \tanh\pi\lambda.$

\end{appendix}


\begin{thebibliography}{999}

\bibitem{BeGa}
Berenstein C.A., Gay R.: {\it Complex Variables}. Springer, New York 1991.

\bibitem{BeKo}
Berezanskii Iu.M., Kondratiev Iu.M.: {\it Spectral Methods in Infinite-Dimensional Analysis}, Vol. {\bf 1}. Kluwer Academic Publishers, Dordrecht 1995

\bibitem{BBMS} Bertola M., Bros J., Moschella U., Schaeffer R.: Decomposing quantum fields on branes. Nuclear physics {\bf B 581}, 575-603 (2000)

\bibitem{Bog} Bogachev V.I.: Gaussian Measures, AMS Providence
1998 (translated from Russian).

\bibitem{BEM} Bros J., Epstein H., Moschella U.: Towards a general theory of quantized fields on the anti-de Sitter spacetime. Commun. Math. Phys. {\bf 231}, 481-528 (2002)

\bibitem{Di}
Dimock J.: Markov quantum fields on a manifold. Rev. Math. Phys. {\bf 16}, 243-256 (2004)

\bibitem{DiGl}
Dimock J., Glimm J.: Measures on Schwartz distribution space and applications to $P(\phi)_2$ field theories. Adv. Math. {\bf 12}, 58-83 (1974)

\bibitem{Do}
Dobrev V. K.: Intertwining operator realization of the AdS/CFT correspondence.  Nuclear Phys. {\bf B 553}, 559--582 (1999)
\bibitem{DuRe}
D\"utsch M., Rehren K.H.: A comment on the dual field in the AdS-CFT correspondence. Lett. Math. Phys. {\bf 62}, 171-184 (2002)

\bibitem{EgOk}
Eguchi M., Okamoto K.: The Fourier transform of the Schwartz space on a symmetric space. Proc. Japan Acad. {\bf 53}, Ser. A, 237-241 (1977)

\bibitem{GeSh}
Gelfand I.M., Shilov G.E.: {\it Generalized Functions}, Vol.{\bf 1}. Properties and Operations. Academic press, New York-London 1964 (1977)

\bibitem{GlJa}
Glimm J., Jaffe A.: {\it Quantum Physics. A Functional Integral Point of View}, 2nd edition. Springer, New-York 1987

\bibitem{Go}
Gottschalk H.: {\it Die Momente gefalteten Gau\ss -Poissonschen
wei\ss en Rauschens als Schwingerfunktionen}. Diploma thesis, Bochum
1995
\bibitem{GuKlPo}

Gubser S.S., Klebanov I.R., Polyakov A.M.: Gauge theory correlators from noncritical
string theory. Phys. Lett. {\bf B 428}, 105-114 (1998)

\bibitem{Ha}
Haba Z.: Quantum field theory on manifolds with a boundary. J. Phys. {\bf A 38}, 10393-10401 (2005)

\bibitem{He}
Helgason S.: {\it Groups and Geometric Analysis. Mathematical Surveys and Monographs}, Vol. {\bf 83}, Providence, RI 2000

\bibitem{JR} Jaffee A., Ritter G.: Quantum field theory on curved
backgrounds II: Spacetime symmetries. arXiv:0704.0052v1 [hep-th]

\bibitem{Kni}
Kniemeyer O.: {\it Untersuchungen am erzeugenden Funktional der AdS-CFT-Korrespon\-denz}. Diploma thesis, Univ. G\"ottingen 2002

\bibitem{LiTs}
Liu Hong, Tseytlin A.A.: On four point functions in the CFT/AdS correspondence.
Phys. Rev. {\bf D 59}:086002 (1999)

\bibitem{Ma}
Maldacena J.: The large N limit of superconformal field theories and supergravity. Adv.
Theor. Math. Phys. {\bf 2}, 231-252 (1998)

\bibitem{muwi}
M\"uck W., Wiswanathan K.S.: Regular and irregular boundary conditions in the AdS/CFT Correspondence. Phys. Rev. {\bf D 60}:081901 (1999)

\bibitem{OsSch1}
Osterwalder K., Schrader R.: Axioms for Euclidean Green's functions.  Comm. Math. Phys. {\bf 31}, 83-112 (1973)

\bibitem{OsSch2}
Osterwalder K., Schrader R.: Axioms for Euclidean Green's functions. II. With an appendix by Stephen Summers.  Comm. Math. Phys. {\bf 42},  281--305 (1975)

\bibitem{Rea}
Rehren K.-H.: Algebraic holography. Ann. Henri Poincar\`{e} {\bf 1}, 607-623 (2000)

\bibitem{Reb}
Rehren, K.-H.: QFT lectures on AdS-CFT, arXiv:hep-th/0411086v1

\bibitem{Str}
Strichartz R. S.: Analysis of the Laplacian on the complete Riemannian manifold.  J. Funct. Anal. {\bf 52}, 48-79 (1983)

\bibitem{tar}
Tartaru D.: Strichartz estimates in the hyperbolic space and global existence for
the nonlinear wave equation. Trans. Am. Math. Soc. {\bf 353},
795-807 (2000)

\bibitem{We}
Weidmann J.: {\it Lineare Operatoren.} B.G. Teubner, Stuttgart 1976

\bibitem{Wi}
Witten E.: Anti-de Sitter space and holography. Adv. Theor. Math. Phys. {\bf 2}, 253-291 (1998)

\end{thebibliography}
\end{document}